\documentclass[11pt]{article}

\usepackage{pb-diagram}
\usepackage{latexsym}
\usepackage{amsfonts}
\usepackage[usenames,dvipsnames]{xcolor}
\usepackage{graphicx,amssymb,amsmath,epsfig}
\usepackage[english]{babel}
\usepackage{graphicx}
\usepackage{dcolumn}
\usepackage{bm}
\usepackage{slashed}

\definecolor{myblue}{rgb}{0,0,0.8}
\usepackage[colorlinks=true
,urlcolor=myblue	
,anchorcolor=myblue
,citecolor=myblue
,filecolor=myblue
,linkcolor=myblue
,menucolor=myblue
,pagecolor=myblue
,linktocpage=true  
]{hyperref}

\catcode`\@=11
\def\marginnote#1{}

\newcount\hour
\newcount\minute
\newtoks\amorpm
\hour=\time\divide\hour by60 \minute=\time{\multiply\hour by60
\global\advance\minute by-\hour}\edef\standardtime{{\ifnum\hour<12
\global\amorpm={am}%
        \else\global\amorpm={pm}\advance\hour by-12 \fi
        \ifnum\hour=0 \hour=12 \fi
        \number\hour:\ifnum\minute<10
0\fi\number\minute\the\amorpm}}
\edef\militarytime{\number\hour:\ifnum\minute<10 0\fi\number\minute}

\def\draftlabel#1{{\@bsphack\if@filesw {\let\thepage\relax
   \xdef\@gtempa{\write\@auxout{\string
      \newlabel{#1}{{\@currentlabel}{\thepage}}}}}\@gtempa
   \if@nobreak \ifvmode\nobreak\fi\fi\fi\@esphack}
        \gdef\@eqnlabel{#1}}
\def\@eqnlabel{}
\def\@vacuum{}
\def\draftmarginnote#1{\marginpar{\raggedright\scriptsize\tt#1}}
\def\draft{\oddsidemargin -.5truein
        \def\@oddfoot{\sl preliminary draft \hfil
        \rm\thepage\hfil\sl\today\quad\militarytime}
        \let\@evenfoot\@oddfoot \overfullrule 3pt
        \let\label=\draftlabel
        \let\marginnote=\draftmarginnote

\def\@eqnnum{(\theequation)\rlap{\kern\marginparsep\tt\@eqnlabel}%
\global\let\@eqnlabel\@vacuum}  }

\def\numberbysection{\@addtoreset{equation}{section}
        \def\theequation{\thesection.\arabic{equation}}}

\def\underline#1{\relax\ifmmode\@@underline#1\else
 $\@@underline{\hbox{#1}}$\relax\fi}

\catcode`@=12 \relax

\numberbysection

\topmargin by -\headsep
\newcommand{\bi}{\begin{itemize}}
\newcommand{\ei}{\end{itemize}}
\newcommand{\bc}{\begin{center}}
\newcommand{\ec}{\end{center}}

\newcommand{\be}{\begin{equation}}
\newcommand{\ee}{\end{equation}}

\newcommand{\bqn}{\begin{eqnarray}}
\newcommand{\eqn}{\end{eqnarray}}

\def\br{\begin{eqnarray}}
\def\er{\end{eqnarray}}

\def\({\left(}
\def\){\right)}
\def\[{\left[}
\def\]{\right]}

\def\a{\alpha}
\def\b{\beta}

\def\d{\delta}

\def\bpsi{\bar{\psi}}

\def\eps{\epsilon}

\def\g{\gamma}
\def\G{\Gamma}

\def\o{\over}
\def\om{\omega}

\def\pa{\partial}

\def\s{\sigma}

\def\th{\theta}

\def\vep{\varepsilon}
\def\bvep{\bar{\varepsilon}}

\def\sl{\sqrt{\lambda}}

\def\ba{\begin{align}}
\def\ea{\end{align}}
\def\be{\begin{eqnarray}}
\def\ee{\end{eqnarray}}

\def\a{\alpha}
\def\b{\beta}
\def\g{\gamma}
\def\d{\delta}

\def\s{\sigma}

\def\G{\Gamma}

\def\o{\omega}

\def\bt{\bar{\theta}}
\def\cw{{\cal{W}}}

\def\spa{\slashed{\partial}}

\def\cL{{\cal L}}

\begin{document}

\vspace*{1cm}
\noindent

\vskip 1 cm
\begin{center}
{\Large\bf A supersymmetric exotic field theory in (1+1) dimensions: 
one loop soliton quantum mass corrections}
\end{center}
\normalsize
\vskip 1cm
\begin{center}
{A. R. Aguirre}\footnote{\href{mailto:alexis.roaaguirre@unifei.edu.br}{alexis.roaaguirre@unifei.edu.br}} and  G. Flores-Hidalgo\footnote{\href{mailto:gfloreshidalgo@unifei.edu.br}{gfloreshidalgo@unifei.edu.br}}\\[.7cm]

\par \vskip .1in \noindent
\emph{Instituto de F\'isica e Qu\'imica, Universidade Federal de Itajub\'a\\ 37500-903, Itajub\'a, MG, Brazil.}
\vskip 2cm

\end{center}

\begin{abstract}

We consider one loop quantum corrections to soliton mass for the ${\cal N}=1$ supersymmetric extension of the (1+1)-dimensional scalar field theory with the potential $U(\phi) = \phi^2 \cos^2\left(\ln \phi^2\right)$. First, we compute the one loop  quantum soliton mass correction of the bosonic sector. To do that, we regularize implicitly such quantity by subtracting and adding its corresponding tadpole graph contribution, and use the renormalization prescription that the added term vanishes with the corresponding counterterms. As a result we get a finite unambiguous formula for the soliton quantum mass corrections up to one loop order. Afterwards, the computation for the  supersymmetric case is extended straightforwardly and we obtain
for the one loop quantum correction of the SUSY kink mass the expected value previously derived for the SUSY sine-Gordon and $\phi^4$ models. However, we also have found that for a particular value of the parameters, contrary to what was expected, the introduction of supersymmetry in this  model worsens ultraviolet divergences rather than improving them.

\end{abstract}

\newpage
\tableofcontents

\vskip 1cm
\hspace{-0.73cm}
\rule{16.5cm}{1pt}
\vskip 2.3cm

\section{Introduction}
\label{sec:intro}

The calculations of quantum corrections to the kink mass in (1+1)-dimensional field theories have been an intensively studied subject since many years ago \cite{Dashen,Faddeev}. Originally, authors calculated the quantum corrections to the kink mass in the bosonic $\phi^4$ and sine-Gordon field theories. Some years later, supersymmetric extensions of those models were also studied, and since then a large amount of different approaches to calculate quantum corrections to the supersymmetric kink mass and central charge have been exhaustively investigated \cite{Dadda}--\cite{Graham}. Remarkable efforts were made on dealing with two interesting but tricky issues: whether or not the bosonic and fermionic contributions in the quantum corrections to the supersymmetric kink mass cancel each other, and if the BPS saturation condition survives at quantum level.

After many attempts of solving these two issues without having reached any consensus, it was shown in \cite{Graham}, using a simple renormalization scheme, that the correction to the supersymmetric  kink mass for the $\phi^4$ and sine-Gordon models is given by $\Delta M = -m/2\pi$, which is in complete agreement with some previous results obtained in \cite{Schon,Nastase}. Furthermore, authors also showed in \cite{Graham} that the BPS bound remains saturated at one loop approximation. Soon after, it was obtained the same exact result for the supersymmetric kink mass by using a generalized momentum cut-off regularization scheme \cite{Lit}. 

In all above cited works, authors treated mainly sine-Gordon and $\phi^4$ models because their 
one loop solvability. This is possible since in those cases the kink one loop fluctuations are described by
exactly solvable one dimensional Schr\"odinger  equation corresponding to  the Poschl-Teller type  potentials. Some time ago  it was considered the problem of constructing one loop exactly
solvable two-dimensional scalar models starting from exactly solvable one dimensional Schr\"odinger equations \cite{Gabriel1}.
In particular, from the Scarf II hyperbolic potential, authors obtained an exotic bosonic scalar field model with a potential given by $U(\phi)=\phi^2\cos^2\ln(\phi^2)$ (see also \cite{Lohe,kumar,guilarte}). 
Unlike the sine-Gordon model, this exotic potential exhibits infinite degenerate vacua which are not equivalent, i.e. even thought the second order derivatives of the potential at degenerate minima are equal, higher order derivatives are not. As a consequence, quantum solitons between
adjacent vacua will exist only semiclassically, and then such states will
become unstable at full quantum level as it was already pointed out in the $\phi^6$ model \cite{Rajaraman12}, where authors proposed to couple fermionic fields to the scalar fields in a supersymmetric way to overcome such issue. The quantum instability of the $\phi^6$ solitons has been also discussed more recently in \cite{Weigel2017, Roman2017}.
Therefore, in order to have  meaningful quantum solitons, in this  paper we will consider the
supersymmetric extension of the exotic bosonic scalar field theory.

Specifically, we consider the ${\cal N}=1$ supersymmetric extension of the aforementioned  exotic bosonic scalar field theory and compute the first quantum corrections to the mass of the supersymmetric kink. Since the first quantum mass corrections are in general divergent, we have to deal with
the issue of regularization and renormalization. We will do this task by using a modification  of the scattering phase shift method, which requires the use  of the bosonic and fermionic phase shifts,  and the expression for the quantum mass corrections in terms of the Euclidean effective action. In order to assure the correctness of our method
we first perform the one loop computations for the soliton mass in the purely bosonic sector limit to compare with results previously obtained. Afterwards we extend the method to the  supersymmetric case   for which we have found results that also agree with
the ones previously obtained for different models. However, we have found also an unexpected and curious result. It turns to be that for a particular value of the parameter of the model, the introduction of supersymmetry seems to worse the ultraviolet divergences rather than improving them. 

This paper is organized as follows. In section 2, for the sake of clarity we present some basics on ${\cal N}=1$ supersymmetric field theory and then introduce the supersymmetric extension of the exotic bosonic potential. In section 3, we compute the one loop  kink quantum mass corrections of the bosonic sector. In section 4, we extend the computation for the supersymmetric kink mass. Finally, some concluding remarks are presented in section 5.


\section{The ${\cal N}=1$ supersymmetric field theory}

In this section we will introduce a supersymmetric field theory as an extension of the bosonic scalar model previously studied in \cite{Gabriel1}. In the standard superspace approach, the dynamics of the theory is derived from an action depending on some superfields. These superfields are functions in the superspace, which is constructed by adding a two-component Grassmann variable \mbox{$\theta_\a=\{\th_1,\th_2\}$} to the two-dimensional space-time $x^\mu=\{t,x\}$.  Starting from one real bosonic field $\phi(x)$, we can define a bosonic superfield as follows,
\br
\Phi(x,\th) =\phi(x) + \bt \psi(x) +\frac{1}{2}\bt\th F(x),
\er
where $\psi(x)$ is a two-component Majorana spinor, and $F(x)$ a real auxiliary bosonic field. Here, we have used the usual convention $\bt = \th \g^0$, where the representation of the $\g$-matrices in the two-dimensional space is chosen to be
\br 
\g^0 =\s_2=\left(\begin{array}{cc} 0&-i\\i&0\end{array}\right),\qquad \g^1=i\s_3 =\left(\begin{array}{cc} i&0\\0&-i\end{array}\right), \qquad \g^5 = \g^0\g^1 = \left(\begin{array}{cc} 1&0\\0&-1\end{array}\right).
\er
Under a translation in the superspace,
\br
 x^\mu \to x^\mu -i\bt\g^\mu\vep,  \qquad \th_\a\to\th_\a +\vep_\a,
\er
where $\vep_\a = \{\vep_1,\vep_2\}$ is a constant Grassmannian spinor, the fields transforms as follows,
\br
 \d\phi = \bvep \psi,\qquad  \d\psi = -i\pa_\mu \phi\g^\mu \vep +F \vep,\qquad \d F=-i\bvep \g^\mu \pa_\mu\psi .
 \label{susy}
\er
The most general on-shell action invariant under the SUSY transformations (\ref{susy}) can be written in the following form,
\begin{eqnarray}
 S = && \int d^2x \Big[\frac{1}{2}\(\pa_\mu\phi\pa^\mu\phi + i \bpsi \g^\mu \pa_\mu \psi \) -\frac{1}{2} [\cw'(\phi)]^2 -\frac{1}{2}\cw''(\phi)\bpsi\psi \Big], \label{act2.11}
\end{eqnarray}
where the usual notation $\cw'(\phi) =d\cw(\phi)/d\phi$ has been used. Now, by expanding the above expression around a classical bosonic field configuration $\phi_c$, i.e $\phi=\phi_c+\eta$, up to quadratic order in the fields $\eta$ and $\psi$, we get
\begin{eqnarray}
S&=&\frac{1}{2}\int d^2x \Big[\partial_\mu\phi_c\partial^\mu\phi_c-\big({\cal W}_c'\big)^2+\bar{\psi}\big(i\gamma^\mu\partial_\mu-{\cal W}''_c\big)\psi +\partial_\mu\eta\partial^\mu\eta \nonumber\\
&&~~~~~~\qquad
-\big(({\cal W}'_c)^2+{\cal W}'_c{\cal W}'''_c\big)\eta^2\Big],\quad \mbox{}
\label{expansion}
\end{eqnarray}
where we have denoted ${\cal W}_c \equiv {\cal W}(\phi_c)$, and the field $\phi_c$ satisfies the classical bosonic equation of motion
\begin{equation}
\Box \phi_c+{\cal W}'_c{\cal W}''_c=0.
\label{classical}
\end{equation}
Now, it is well-known that by considering finite energy static kink solutions of Eq. (\ref{classical}), it is possible to obtain the energy of the ground state at one loop order after quantization. Then, as usual after subtracting the
vacuum energy  in the absence of the kink,  we get that the mass of the kink state at one loop order is given by
\begin{equation}
M_{bare}=E[\phi_c]+\frac{1}{2}\sum_{n}\omega_{nb}-\frac{1}{2}\sum_{n}\omega_{nf},
\label{qmass}
\end{equation}
where $E[\phi_c]$ is the energy of the static classical configuration, 
\begin{equation}
E[\phi_c]=\int_{-\infty}^\infty dx \left[\frac{1}{2}\left(\frac{d\phi_c}{dx}\right)^2+\frac{1}{2}[{\cal W}'(\phi_c)]^2\right],
\label{cmass}
\end{equation}
and $\omega_{nb}$ and $\omega_{nf}$ are solutions of the following eigenvalue equations,
\begin{equation}
\left[-\frac{d^2}{dx^2}+({\cal W}'_c)^2+{\cal W}'_c{\cal W}'''_c\right]\eta_n=\omega_{nb}^2\,\eta_n, \label{bosef}
\end{equation}
and
\begin{equation}
\left[-i\g^5\frac{d}{dx}+{\cal W}''_c\g^0\right]\psi_n=\omega_{nf}\,\psi_n.
\label{fermif}
\end{equation}

\begin{figure}
\begin{center}
 \includegraphics[width=8cm,height=5cm]{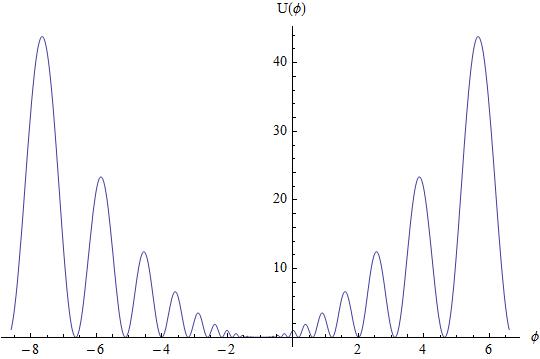}
 \caption{\small Plot of the bosonic potential $U(\phi)$ for $\b=1$ and $B=0.1$}\label{potU}
 \end{center}
\end{figure}

Let us now consider the following  form for the superpotential, namely
\begin{eqnarray}
\cw(\phi) =&&\frac{8m^3B}{\lambda (1+4B^2)}(1+\beta\phi)^2
 \left[\cos\left(\frac{ \ln (1+\beta\phi)^2}{2B}\) +\frac{1}{2B}
 \sin\left(\frac{ \ln\big(1+\beta \phi)^2}{2B}\)\right], \quad \mbox{}
\label{superpotential}
\end{eqnarray}
as the natural supersymmetric extension of the aforementioned exotic bosonic scalar potential, where $m$ is a mass parameter, and  the real parameters $B$, $\lambda$, and $\b$ satisfy the relation $\beta={\sqrt{\lambda}B}/{2m}$. Note that, the superpotential (\ref{superpotential}) becomes the ${\cal N}=1$ super sine-Gordon model superpotential when $B\to 0$.
After substituting in the action (\ref{act2.11}), we get the Lagrangian density
\br 
{\cal L}=&& \frac{1}{2}(\pa_\mu\phi)(\pa^\mu\phi) -\frac{2m^4}{\lambda}(1+\beta  \phi)^2\cos^2\left(\frac{ \ln (1+\beta\phi)^2}{2B}\)+\frac{i}{2} \bpsi \g^\mu \pa_\mu \psi 
\nonumber\\
&&~~~~~~
-\frac{m}{2}\left[B\cos\left(\frac{ \ln (1+\beta \phi)^2}{2B}\) -
\sin\left(\frac{\ln (1+\beta \phi)^2}{2B}\)\right]\bpsi\psi . \label{Lagsusy}
\er
As it was noted in \cite{Gabriel1}, the bosonic potential in the Lagrangian (\ref{Lagsusy}),
\br
 U(\phi) = \frac{1}{2} \({\cal W}'(\phi)\)^2 = \frac{2m^4}{\lambda}(1+\beta  \phi)^2\cos^2\left(\frac{ \ln (1+\beta\phi)^2}{2B}\), \label{bospot}
\er
has infinitely degenerate trivial vacua at the points $\phi_n$ given by
\br
\phi_n=\pm  \frac{2m}{B\sqrt{\lambda}} \left\{\exp\[\left(n+\frac{1}{2}\right)\pi B\right]-1\right\},
\er
where $n=0,\pm 1, \pm 2 ,...$. It can be seen in Figure \ref{potU} that this potential possesses a reflection symmetry around the point $\phi=-1/\b$. 
In addition, we note that $\cw''(\phi_n)$ is equal to $-m$ for $n$ even, and $+m$ when $n$ is odd. This fact implies that the curvature of the potential $U''(\phi_n)$ is the same for all $n$, namely,
\br
 U''(\phi_n) = m^2, \qquad  n=0,\pm 1, \pm 2, ...
 \label{m2}
\er
The classical bosonic kink and anti-kink solutions have the following explicit form,
\begin{eqnarray}
 \phi_c(x) &=& \pm \frac{2m}{B\sqrt{\lambda}}\left\{
  \exp\left[B\(n\pi + \eps \,\tan^{-1}\sinh(mx)\)\]-1\right\},
\label{gkink}
\end{eqnarray}
where $\eps =+1$ corresponds to a kink solution, while \mbox{$\eps=-1$} corresponds to the anti-kink solution. The corresponding classical masses of the kink (or anti-kink) are given by,
\begin{eqnarray}
 E_n[\phi_c] &=& |{\cal W}(\phi_{n+1})-{\cal W}(\phi_n)|\nonumber \\ &=& E_0 \exp\(2n\pi B\),\, \label{masses}
\er
where 
\br
 E_0 =  \frac{8m^3} {\lambda(1+4B^2)}\cosh\(\pi B\). \label{Clasmass}
\er
Despite of the exponential dependence in eq. (\ref{masses}), the classical masses for a kink (or anti-kink) connecting any two neighbouring vacua is finite, and satisfy the relation \mbox{$(E_{(n\pm 1)}/E_n )= e^{\pm 2\pi B}$.} It is also worth pointing out that in the limit $B\to0$ we recover the kink configuration for the sine-Gordon model from  eq. (\ref{gkink}), as well as its corresponding classical mass from eq. (\ref{Clasmass}). 
%


\section{Bosonic one loop quantum mass corrections}
Let us consider first the purely bosonic case described by the following Lagrangian density,
\begin{equation}
{\cal L}=\frac{1}{2}\partial_\mu\phi\partial^\mu \phi-U(\phi) +\d \cL,
\label{redefined}
\end{equation}
where the potential $U(\phi)$ is given by Eq. (\ref{bospot}), and $\d\cL$ contains adequate counterterms in order to render finite the theory. By quantizing around the static kink $\phi_c$, we get for the soliton mass at one loop order,
\begin{equation}
M_b=E[\phi_c]+ \d M_b +\frac{1}{2}\sum_{n}\omega_{nb}
-\frac{1}{2}\sum_{k}\omega_{b}^0(k)\;,
\label{e1}
\end{equation}
where the index $b$ stands for bosonic contributions,  $\d M_b$ are the counterterm contributions from the $\d\cL$ term. For simplicity, eq. (\ref{e1}) can be written in the following way,
\br
 \Delta M_b = \Delta M_+ + \d M_b,
\er
where $\Delta M_b = M_b - E[\phi_c]$, and
\br
\Delta M_+ = \frac{1}{2}\sum_{n}\omega_{nb}
-\frac{1}{2}\sum_{k}\omega_{b}^0(k). \label{deltamp}
\er
The eigenfrequencies $\omega_{nb}$ are given by Eq. (\ref{bosef}), which can be rewritten as 
\begin{equation}
\left[-\frac{d^2}{dx^2}+V_{+}(x)\right]\eta_n=\omega_{nb}^2\eta_n,
\label{bosef1}
\end{equation}
with
\begin{equation}
V_{+}(x)=U''[\phi_c(x)]\,=\, ({\cal W}_c'')^2+{\cal W}_c' {\cal W}_c'''.\label{equ3.4}
\end{equation}
Also, the free soliton eigenfrequencies $\omega_{b}^0(k)$ are given by
\begin{equation}
\left[-\frac{d^2}{dx^2}+m^2 \right]\eta_k=[\omega_{b}^0(k)]^2\eta_k,
\label{freeboson}
\end{equation}
where $m$ represents the mass of the quantum fluctuations
around the trivial vacua. The term $\Delta M_+$ in eq. (\ref{deltamp}) is  logarithmically divergent, and there are several techniques to deal with
this issue in the literature \cite{chan,dunne,Graham98,spectral,Goldhaber,Bordag,Parnachev, Wimmer}. Here, we will consider a simple
method to regularize that term based on the following formal identity \cite{boya},
{
\begin{eqnarray}
\Delta M_+  
&=&\frac{1}{2}\int_{-\infty}^\infty\frac{d\omega}{2\pi}\,{\rm tr}\ln
\left(1+\hat{A}_{+}\right) =\hspace{0.2cm}\parbox[b]{7.5cm}{ \raisebox{-0.9999cm}
{\psfig{file=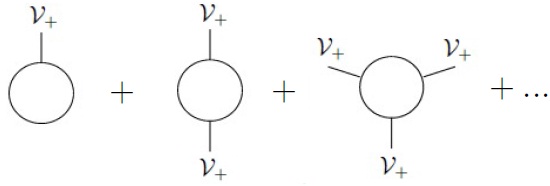,height=2.3cm,width=6cm}}}
\label{eaction1}
\end{eqnarray}}
where 
\begin{equation}
\hat{A}_{+}=\frac{{\cal V}_+}{\omega^2-\frac{d^2}{dx^2}+m^2},
\label{operator}
\end{equation}
and ${\cal V}_+=V_+(x)-m^2$. Equation (\ref{eaction1}) is the one loop quantum correction to the kink
mass expressed as the Euclidean effective action per unit time. From its expansion in
terms of Feynman graphs, we identify the following tadpole graph contribution, 
\begin{equation}
\parbox[b]{2cm}{ \raisebox{-0.3cm}{\psfig{file=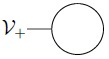,
height=0.85cm,width=1.55cm}}}\quad\hspace{-0.7cm}
= \frac{1}{2}\!\int_{-\infty}^\infty\! \frac{d\omega}{2\pi} \,{\rm tr}\hat{A}_{+}\;.
\label{tadpole}
\end{equation} 
as the only ultraviolet divergent graph. Therefore, by adding and subtracting  the above tadpole graph  in (\ref{e1}), and using the renormalization prescription
that the added tadpole graph cancels with $\delta M_b$, we will get a 
finite result for the one loop soliton mass,
\begin{eqnarray}
\Delta M_b= \Delta M_+
- \frac{1}{2}\int_{-\infty}^\infty\! \frac{d\omega}{2\pi} {\rm tr}\hat{A}_{+}.\quad\mbox{}
\label{renormalized}
\end{eqnarray}
Of course, each one of the  terms in eq. (\ref{renormalized}) is separately divergent, but their difference is not. Therefore, if the same scheme is used to compute them, we must get a finite unambiguous result  for $M_b$, independently of the regularization scheme used. Using the phase shift method \cite{Graham98,spectral}, it is not difficult to get for the one loop
soliton quantum mass correction the following result\footnote{See for instance Eq. (1.17) in \cite{spectral}.},
\bqn
\Delta M_{b}&=&\frac{1}{2}\sum_{i=1}^{N}\(\omega_{ib}-m\)
-
\frac{1}{2\pi}
\int_{0}^{\infty}\frac{k}{\omega_b(k)}\left[\delta_+(k) +
\frac{\langle {\cal V}_+\rangle}{2k}\right]dk,\quad\mbox{}
\label{unloop}
\eqn
where 
\begin{equation}
\langle {\cal V}_+\rangle=\int_{-\infty}^\infty dx \,\,{\cal V}_+(x),
\label{born}
\end{equation}
$N$ is the number of discrete eigenfrequencies $\omega_{ib}$, and the phase shift
$\delta_+(k)$ can be obtained directly from the scattering $S$ matrix as
\begin{eqnarray}
\delta_+(k)&=&\frac{1}{2i}\ln\det S(k)\nonumber\\&=& \frac{1}{2i}\ln\left[\frac{T(k)}{T^\ast(k)}\right],\label{11}
\end{eqnarray}
where in passing to the second line we have used \cite{Bianchi}
\begin{equation}
S = 
\begin{bmatrix}
    T(k)  & -T(k)R^\ast(k)/T^\ast(k) \\
    R(k)  & T(k)
\end{bmatrix}.
\label{resp2}
\end{equation}
Here  $R$ and $T$ denote respectively the reflection and transmission coefficients of the one
dimensional scattering problem described by the continuous spectrum  of (\ref{bosef1}).

Let us now apply the general formula (\ref{unloop}) for the bosonic density potential (\ref{bospot}). In this case, by substituting the static configuration (\ref{gkink}) in eq. (\ref{equ3.4}), we get the potential
\br 
 V_+(x) = m^2\left[1+\frac{(B^2-2)}{\cosh^2(mx)}-3B\frac{\sinh(mx)}{\cosh^2(mx)} \], \label{potmais}
\er
which belong to the Scarf II hyperbolic exactly solvable potentials \cite{Cooper}. It is worth pointing out that this potential does not depend on the index $n$ of the static field configuration $\phi_c$.  This potential is shown in Figure \ref{Vmais}. It has only one discrete eigenvalue, namely the zero mode \mbox{$\omega_{0b}=0$}, and its corresponding eigenfunction has the following form,
 \br 
 \eta_{0}(x) &=&\frac{c_0}{\cosh(mx)}\,{\exp\left[B\tan^{-1}(\sinh(mx))\]},
\er
with $c_0$ a normalization constant.
\begin{figure}
\begin{center}
 \includegraphics[width=7cm,height=5.5cm]{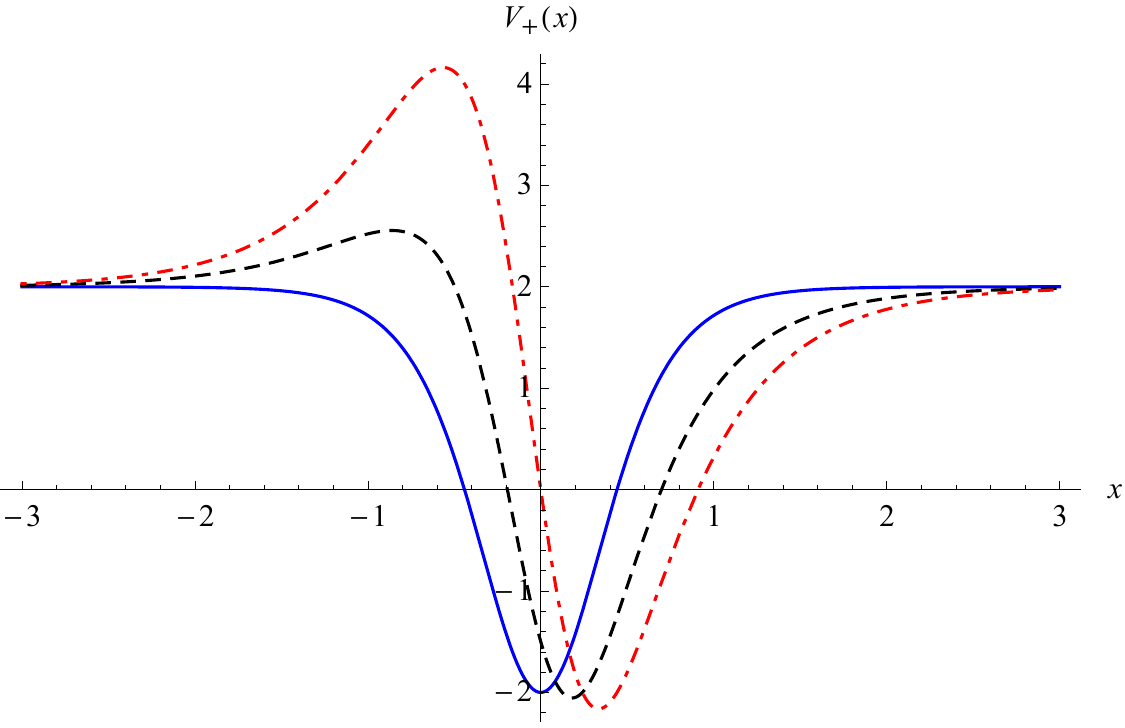}
 \caption{\small Plot of the potential  $V_+(x)$ for $B=0$ (solid line), $B=0.5$ (dashed line), and $B=1$ (dotdashed line), respectively}\label{Vmais}
\end{center}
\end{figure}
In addition,  the transmission coefficient amplitude for this potential is given by \cite{Cooper},
\br 
 T_+ (k)&=& \frac{(-1 +\frac{ik}{m})\G\big(\frac{1}{2}- iB- \frac{ik}{m}\big)\G\big(\frac{1}{2}+ 
 i B-\frac{ik}{m}\big)}{(1+\frac{ik}{m})\G^2\big(\frac{1}{2}-\frac{ik}{m}\big)}.\qquad\,\,\,\, \mbox{}\label{tmais}
\er
Now, by substituting Eqs. (\ref{potmais}) and (\ref{tmais}) in Eqs. (\ref{born}) and (\ref{11})  respectively, we find that 
\be
 \mbox{$
\langle{\cal  V}_+\rangle = 2m (B^2-2)$},
\label{meanvalue}
\ee
and the bosonic phase shift is given by
\br 
 \delta_+(k) = 2\arctan\Big(\frac{m}{k}\Big)\! +\frac{1}{2i}\ln\!\left[\frac{\G\big(\frac{1}{2}+iB-\frac{ik}{m}\big)}
{\G\big(\frac{1}{2}+iB +\frac{ik}{m}\big)}\right]\!\!
+\frac{1}{2i}\ln\!\left[\frac{ 
 \G\big(\frac{1}{2}-iB -\frac{ik}{m}\big)\G^2\big(\frac{1}{2}+\frac{ik}{m}\big)}
 {\G\big(\frac{1}{2}-iB+\frac{ik}{m}\big)\G^2\big(\frac{1}{2}-\frac{ik}{m}\big)}\]\!\!,\quad\,\,\,\,\,\mbox{}\label{equ3.29}
\er
which is plotted in Figure \ref{fbos} for different values of the parameter $B$. It can be verified from eq. (\ref{equ3.29}) that $\delta_+(0)=\pi$ and $\delta_+(+\infty)=0$, which is consistent with the Levinson theorem,  where there is one half-bound  state. This can be seen by noting that the transmission coefficient does not vanish at the threshold $k=0$, or from the graph of the phase shift plotted in Figure \ref{fbos}.
\begin{figure}
\begin{center}
 \includegraphics[width=7cm,height=5.5cm]{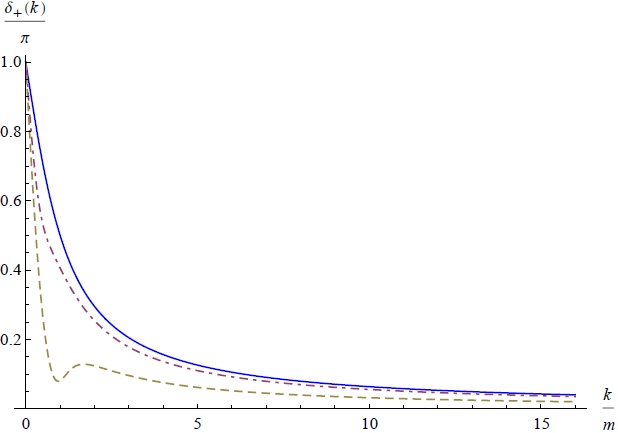}
 \caption{\small Plot of the bosonic phase phifts $\d_+(k)/\pi$ as a function of $k/m$. The solid, dotdashed and dashed lines correspond to $B=0$, $B=0.5$, and $B=1$, respectively}\label{fbos}
\end{center}
\end{figure}
Now, by using the above results for the potential and the phase shift, and substituting in eq. (\ref{unloop}), we finally get
\br
 \frac{\Delta M_b}{m} &=& \frac{(B^2-2)}{2\pi}-\int_{0}^{\infty}\frac{dk}{2\pi}\frac{B^2}{\sqrt{1+k^2}}\nonumber\\
 & &+
 \int_{0}^{\infty}\frac{dk}{4i\pi }\sqrt{1+k^2}\frac{d}{dk}\ln\left[\frac{\G\big(\frac{1}{2}+iB-ik\big) \G\big(\frac{1}{2}-iB -ik\big)\G^2\big(\frac{1}{2}+ik\big)}{\G\big(\frac{1}{2}+iB 
 +ik\big)\G\big(\frac{1}{2}-iB+ik\big)\G^2\big(\frac{1}{2}-ik\big)} 
\right].\label{correcao}\qquad \mbox{}
\er
Taking $B=0$ in above expression, the integral vanishes and we get $\Delta M_b=-m/\pi$, the well-known value for one
loop quantum corrections for the soliton mass in sine-Gordon model. This was expected since as  already mentioned, in
the limit $B\to 0$, the density potential (\ref{bospot}) reduces to sine-Gordon one.  For other values of $B$ it is not possible
to compute the integral in (\ref{correcao}) analytically, however we can integrate  numerically straightforwardly by using for example Mathematica software. In Figure \ref{delta3}, we have plotted the results for $\Delta M_b/m$ as function of parameter $B$, ranging from $B=0$ to
$B=3.2$, where we have restricted to $B\geq 0$ because $\Delta M_b(-B)=\Delta M_b(B)$ as can be concluded from (\ref{correcao}).
For  $B>3.2$, the behaviour of $\Delta M_b(B)$ is quite similar to the one depicted in Figure \ref{delta3},  it varies smoothly as a function of
parameter $B$ and  $\Delta M_b(B+h)<\Delta M_b(B)$, for sufficiently large positive $h$.
\begin{figure}
\begin{center}
 \includegraphics[width=8cm,height=5cm]{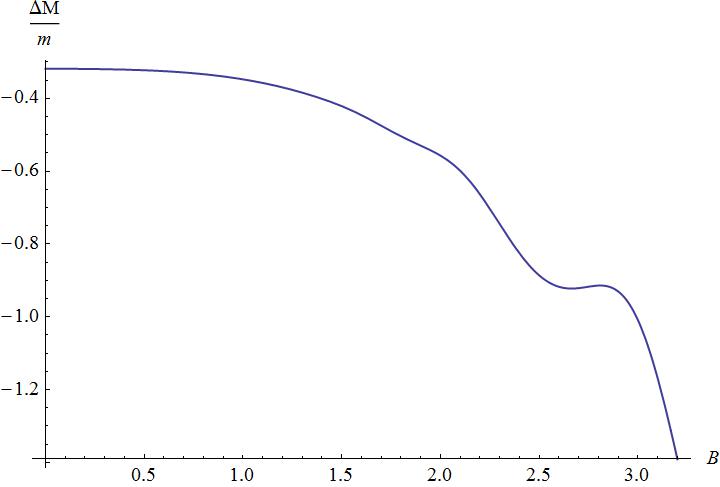}
 \caption{\small Plot of $\Delta M_b/m$ as a function of the parameter $B$}\label{delta3}
 \end{center}
\end{figure}


\newpage
\section{One loop SUSY kink quantum mass correction}
In this section, we compute the quantum correction to the SUSY kink mass by including both bosonic and fermionic fluctuations. First of all, let us now consider fluctuations of the fermion field $\psi(x)$. The Fermi field equation is given by,
\br
 [i\spa -\cw''(\phi)]\psi = 0.
\er
From the kink configuration of the bosonic field $\phi_c(x)$ given in (\ref{gkink}), we can write the fermionic fluctuations in the following form,
\br
 \psi (x,t) = \xi(x) e^{-i\o_f t} + \xi^*(x)e^{i\o_f t} ,
\er
with 
\br
 \xi(x) = \left(\begin{array}{c} \xi_-(x) \\ \xi_+(x) 
 \end{array}\),
\er
where $\om_f$ is a real variable, and $\xi_\pm(x)$ are static normalizable solutions of the following  system,
\br
 \(\frac{d}{dx} + \cw_c''\)\xi_- &=& -i\o_f \,\xi_+, \nonumber\\[0.1cm]
 \(\frac{d}{dx} - \cw_c''\)\xi_+ &=& -i\o_f \,\xi_-, \label{sys4.5}
\er
Then, by cross-differentiating the system  (\ref{sys4.5}) we obtain the decoupled equations,
\br
 \left\{\frac{d^2}{dx^2} + \left[\o_f^2 - (\cw_c'')^2 +\cw_c'''\,\cw_c'\]\right\} \xi_-(x) &=& 0, \label{ed1}\\
 \left\{\frac{d^2}{dx^2} + \left[\o_f^2 -  (\cw_c'')^2 -\cw_c'''\,\cw_c'\]\right\} \xi_+(x) &=& 0.\label{ed2}
\er
We note that eq. (\ref{ed2}) is the same bosonic fluctuation equation. The system of equations (\ref{ed1}) and (\ref{ed2}) can be rewritten in the following form,
\br
 \left\{-\frac{d^2}{dx^2} + \left[V_\pm(x)-\o_f^2 \]\right\} \xi_\pm(x) &=& 0, 
 \label{added1}
\er
with
\br
 V_\pm(x) &=& (\cw_c'')^2 \pm\cw_c'''\,\cw_c'.\label{Vmenos}
\er
Now, the  one loop quantum mass corrections to the SUSY kink will be given by
\begin{eqnarray}
 \Delta M&=&M-E[\phi_c]\nonumber\\
&=& \frac{1}{2}\sum_{n}\omega_{nb}-\frac{1}{2}\sum_{n} \omega_{nf}+\delta M,
\label{susyma}
\end{eqnarray}
where $\delta M$ is a supersymmetric counterterm, and
$\omega_{nb}$ and $\omega_{nf}$ are  respectively the bosonic and fermionic eigenfrequencies of Eqs. (\ref{bosef1}) and (\ref{added1}). Since the free bosonic and fermionic eigenfrequencies are the same,  we can rewrite eq. (\ref{susyma}) as,
\begin{equation}
\Delta M=\Delta M_+-\Delta {M}_f+\delta M
\label{susymass}
\end{equation}
where $\Delta M_+$ is given in eq. (\ref{deltamp}), and 
\begin{equation}
\Delta M_f=\frac{1}{2}\sum_n\omega_{nf}-\frac{1}{2}\sum_k\omega_f^{0}(k).
\label{fermi1}
\end{equation}
On the other hand,  the fermionic eigenfrequencies are given by eqs. (\ref{ed1}) or (\ref{ed2}). Therefore eq. (\ref{fermi1}) can be rewritten as follows,
\begin{equation}
\Delta M_f=\frac{1}{2}\left(\Delta M_+ +\Delta M_-\right),
\label{fermi2}
\end{equation}
where $\Delta M_-$ is given by an expression similar to (\ref{deltamp}), but now in terms of the
eigenfrequencies described by  Eq. (\ref{ed1}).  In this way, we can rewrite (\ref{susymass}) as
\begin{equation}
\Delta M=\frac{1}{2}\Delta M_{+}-\frac{1}{2}\Delta M_{-}+\delta M.
\label{DeltaM}
\end{equation}
In order to regularize and renormalize the above expression we proceed as in the 
purely bosonic case. Using the formal identity,
\begin{eqnarray}
\frac{1}{2}\sum_{n}\omega_{nb}-\frac{1}{2}\sum_{n} \omega_{nf}&=&\frac{1}{2}\int_{-\infty}^\infty \frac{d\omega}{2\pi}
{\rm tr}\ln\left(\omega^2-\frac{d^2}{dx^2}+V_{+}(x)\right)\nonumber\\
&&-\frac{1}{2}\int_{-\infty}^\infty \frac{d\omega}{2\pi}
{\rm Tr}\ln\left(-i\gamma^0\omega+i\gamma^1\frac{d}{dx}-{\cal W}''_c\right).
\er
Computing the partial trace in the second term of the right-hand side, we get
\br
\frac{1}{2}\sum_{n}\omega_{nb}-\frac{1}{2}\sum_{n} \omega_{nf}&=&\frac{1}{4}\int_{-\infty}^\infty \frac{d\omega}{2\pi}
{\rm tr}\ln\left(\frac{\omega^2-\frac{d^2}{dx^2}+V_{+}(x)}
{\omega^2-\frac{d^2}{dx^2}+V_{-}(x)}\right),
\label{formalidentity}
\end{eqnarray}
and then by expanding the logarithmic term, we find
\begin{eqnarray}
\frac{1}{2}\sum_{n}\omega_{nb}-\frac{1}{2}\sum_{n} \omega_{nf}=&&
\frac{1}{4}\int_{-\infty}^\infty \frac{d\omega}{2\pi} \left(
{\rm tr}\hat{A}_{+}-{\rm tr}\hat{A}_{-}\right)+{\rm finite~terms},\label{susycounterterm}
 \end{eqnarray}
where $\hat{A}_{-}$ is given by an expression similar to (\ref{operator}). The first term
in above expression, is the ultraviolet divergent supersymmetric  tadpole graph.
Now,  adding
and subtracting this term in (\ref{DeltaM}), and using the renormalization
prescription that the added tadpole graph cancels with $\delta M$, we get
the finite result,
\begin{eqnarray}
\Delta M
= &&\frac{1}{2}\left(\Delta M_{+}-\frac{1}{2}\int_{-\infty}^\infty \frac{d\omega}{2\pi} 
{\rm tr}\hat{A}_{+}\right)-\frac{1}{2}\left(\Delta M_{-}-\frac{1}{2}\int_{-\infty}^\infty \frac{d\omega}{2\pi}
{\rm tr}\hat{A}_{-}\right).
\label{susyfinite}
\end{eqnarray}
Note that each term in parentheses of   above expression is of the type (\ref{renormalized}) encountered for the finite result in the purely bosonic case.  Therefore, using (\ref{unloop}),  we get that the 
renormalized one loop correction of the SUSY kink mass is given by
\bqn
\Delta M&=&-\frac{m}{4} -\frac{1}{4\pi}\int_{0}^{\infty} \frac{k}{\omega(k)}\left[
 \big(\delta_{+}(k)-\delta_{-}(k)\big) +
 \frac{\langle  {\cal V}_{+}- {\cal V}_{-}\rangle }{2k}\right]dk.
 \label{renorsusy} 
\eqn
Let us now apply the above formula for the susy exotic field theory. By substituting explicitly the kink configuration (\ref{gkink}) in Eq. (\ref{Vmenos}), we find  that the potential for the upper component $V_-(x)$ takes the following form,
\br
 V_-(x) &=& m^2\left[1 + \frac{B^2}{\cosh^2(mx)}-\frac{B\sinh(mx)}{\cosh^2(mx)}\],\label{pot}
\er
which correspond to the superpartner potential of $V_+(x)$, and also belong to the Scarf II hyperbolic exactly solvable potentials \cite{Cooper}. It has been plotted in Figure \ref{potvmenos}. 
\begin{figure}
\begin{center}
 \includegraphics[width=8cm,height=6cm]{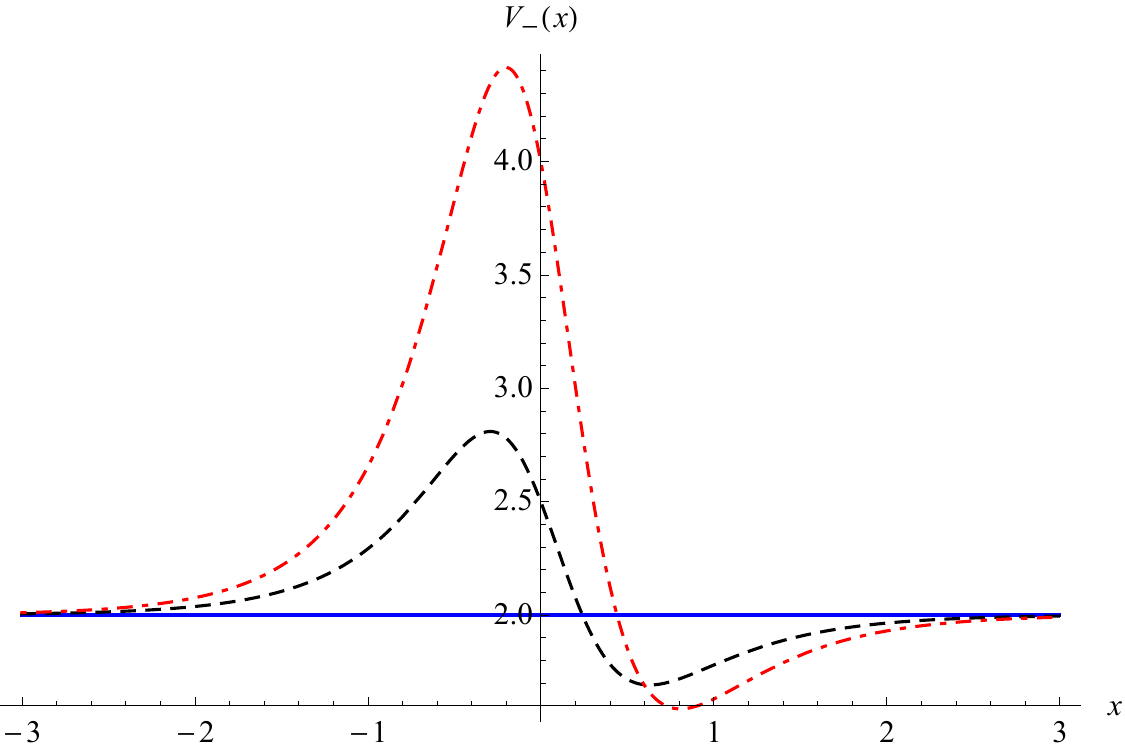}
 \caption{\small Plot of  $V_-(x)$ for $B=0$ (solid line), $B=0.5$ (dashed line), and $B=1.0$ (dotdashed line), respectively}\label{potvmenos}
\end{center}
\end{figure}
The transmission coefficient amplitude for the lower component potential $V_+$ is given by (\ref{tmais}), and for the potential (\ref{pot}) has the following form \cite{Khare88},
\br 
 T_-(k)&=& \frac{\G\big(\frac{1}{2}+ iB-\frac{i k}{m}\big)\G\big(\frac{1}{2}- i B-\frac{i k}{m}\big)}{\G^2\big(\frac{1}{2}-\frac{i k}{m}\big)}.\qquad\,\,\,\, \mbox{}
\er
From above results we find that the phase shift $\delta_+(k)$ is given by (\ref{equ3.29}), whereas the phase shift for the upper component can be written explicitly as follows,
\br 
 \d_-(k)&=&\frac{1}{2i}\ln\left[\frac{\G\big(\frac{1}{2}+iB-\frac{i k}{m}\big)\G\big(\frac{1}{2}-iB -\frac{i k}{m}\big)\G^2\big(\frac{1}{2}+\frac{i k}{m}\big)}{\G\big(\frac{1}{2}+iB +\frac{i k}{m}\big)\G\big(\frac{1}{2}-iB+\frac{i k}{m}\big)\G^2\big(\frac{1}{2}-\frac{i k}{m}\big)}\right].\label{deltamenos}
\er
By comparing Eqs. (\ref{equ3.29}) and (\ref{deltamenos}), we conclude that the phase shifts $\delta_\pm(k)$ satisfy an important relation, namely
\br 
 \d_+(k)- \d_-(k)&=& 2 \arctan\left(\frac{m}{k}\).
\label{deltashift}
\er
Also, from eqs. (\ref{potmais}) and (\ref{pot}) we get,
\begin{equation}
\langle {\cal V}_+-{\cal V}_-\rangle=-4m.
\label{diffv}
\end{equation}
\begin{figure}[t]
\includegraphics[width=7cm,height=5.5cm]{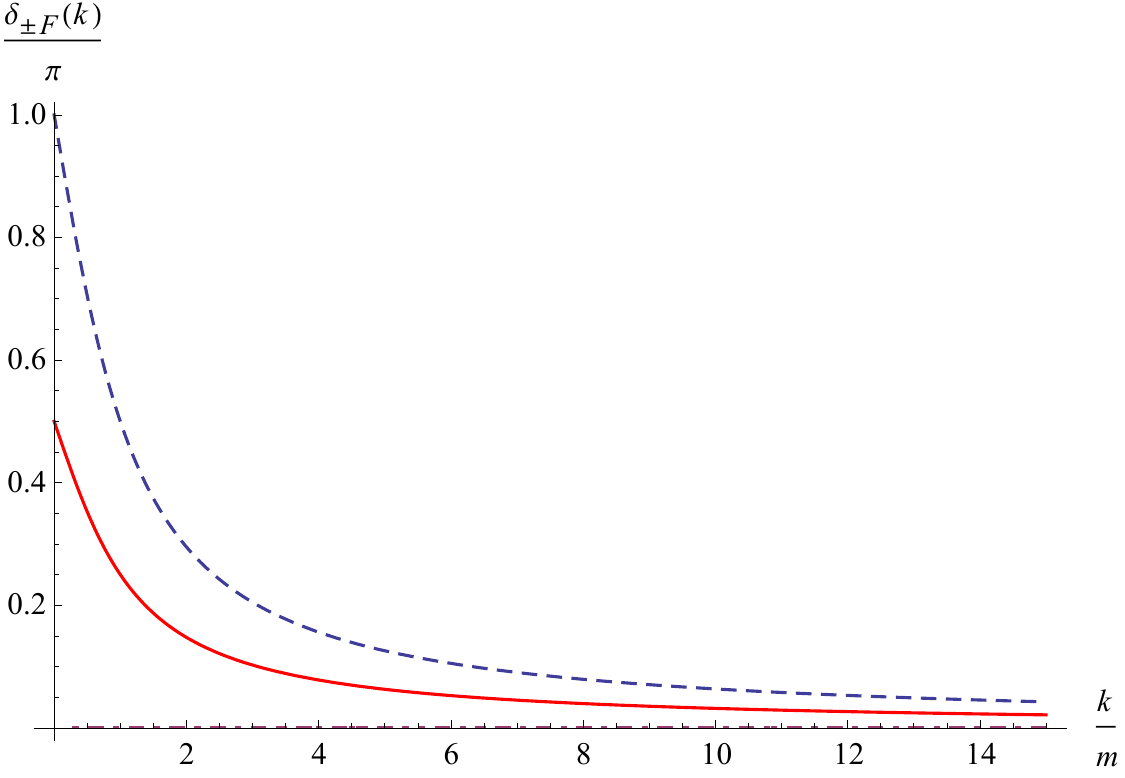}\hspace{0.8cm}\includegraphics[width=7cm,height=5.5cm]{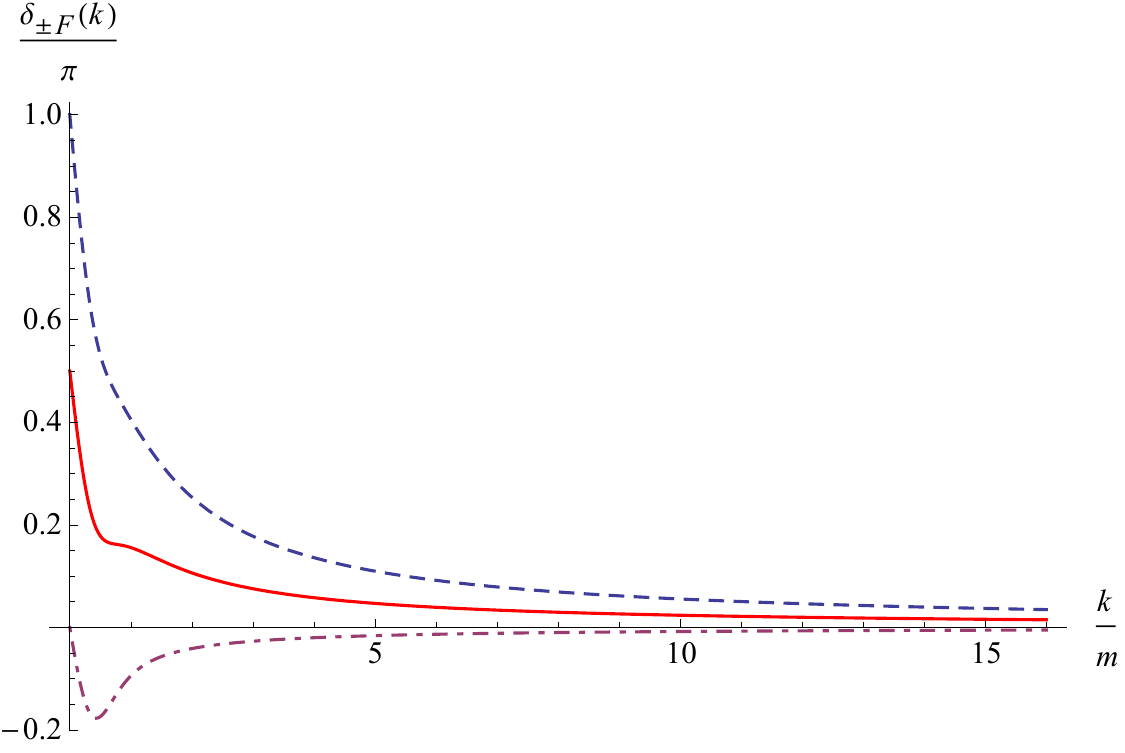}
\begin{center}
 \includegraphics[width=7cm,height=5.5cm]{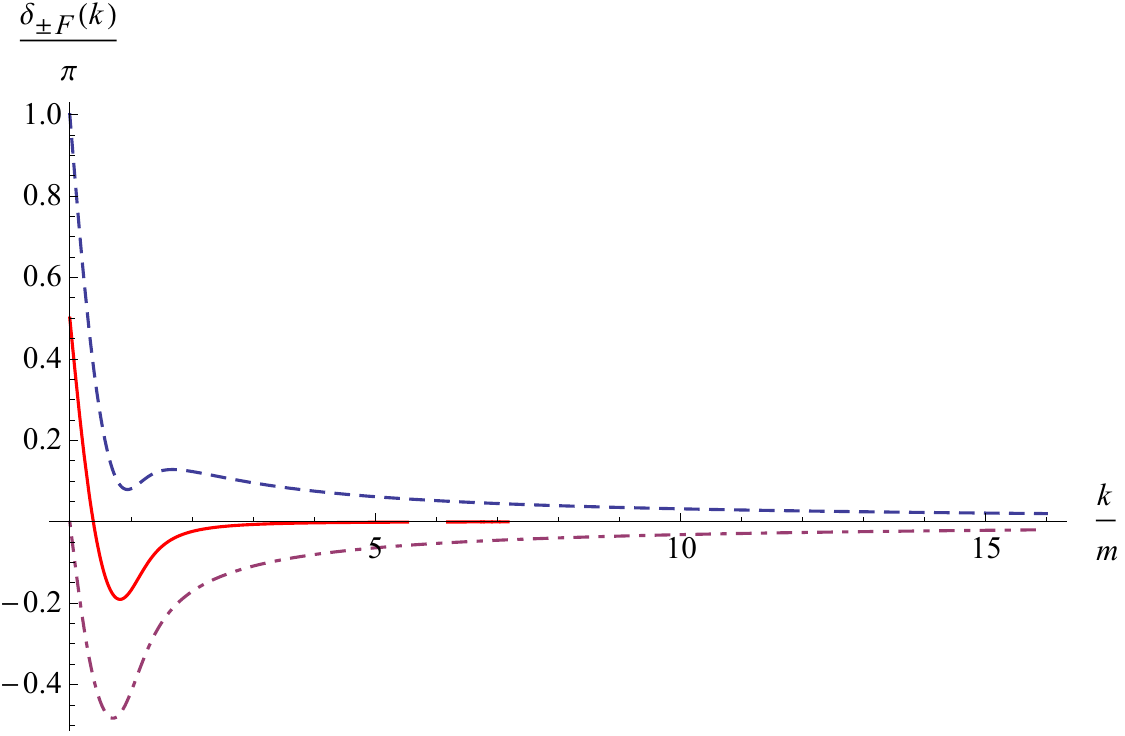}
 \caption{\small Plots of the  phase shifts $\d_{\pm}(k)/\pi$, and $\d_F(k)/\pi$ as functions of $k/m$, for (a) $B=0$, (b) $B=0.5$, and (c) $B=1.0$ . The dashed, solid, and dotdashed lines correspond to $\d_+(k)/\pi$, $\d_F(k)/\pi$, and $\d_-(k)/\pi$, respectively}\label{Figura4}
\end{center}
\end{figure}
\noindent By substituting the above results in Eq.(\ref{renorsusy}), we finally obtain
\begin{equation}
\Delta M=-\frac{m}{2\pi},
\label{finalres}
\end{equation}
which corresponds to the accepted value of the one loop supersymmetric quantum mass correction for any antisymmetric soliton \cite{Graham}.

It is important to note that a fermionic phase shift defined as the average of its values for the upper and lower components, namely
\br
 \d_F(k) &=& \frac{1}{2}\big[\d_+(k)+\d_-(k)\big]\nonumber\\
 &=&\arctan\left(\frac{m}{k}\)+ \d_-(k),
 \label{phasedif}
\er
satisfies the Levinson theorem for $(1+1)$ dimensional Dirac equation \cite{Zhong}, since we have only one bound state corresponding to the zero mode. This definition is consistent with the relation (\ref{fermi2}), and has been also checked previously for several models by using the Levinson theorem \cite{Gou1}--\cite{Gou3}. In Figure \ref{Figura4}, we have plotted these fermionic phase shifts for different values of the \mbox{parameter $B$.}


\section{Concluding remarks}
\label{final}

In this paper we have computed the one loop quantum correction to the kink mass for an exotic supersymmetric theory described by the density Lagrangian (\ref{Lagsusy}) in (1+1) dimensions. For this purpose, we have used a simple regularization scheme based on the formal
identity, (\ref{eaction1}) or (\ref{formalidentity}), and showed that the quantum correction to the mass of the supersymmetric kink up to one loop order is given by $\Delta M =-m/2\pi$, which is in complete agreement with the results reported in the literature \cite{Graham}.

First of all, we have established a formula for computing these corrections in the purely bosonic sector, and then the method has been extended directly for the supersymmetric case. 
In the purely bosonic sector,  we have found that the quantum correction of the kink mass in the limit $B\to0$ is in fact the one of the sine-Gordon model, i.e. $\Delta M_b(B=0) = -m/\pi$. In general, $\Delta M_b(B)$ behaves  as an almost decreasing function for $B>0$ as it can be seen from Figure \ref{delta3}, which shows a smooth interpolating behaviour of $\Delta M_b(B)$ from the sine-Gordon model ($B=0$) to the exotic model ($B>0$). However, after considering the corresponding contributions of the fermionic fluctuations for the quantum corrections we find that the dependence on the parameter $B$ vanishes completely, as well as the divergent part of the SUSY kink mass by adding the appropriate supersymmetric counterterms. 

We would like to call attention to the following curious and interesting result. In the bosonic sector  there
is a special parameter value $B=\sqrt{2}$,  for which the tadpole graph given in (\ref{tadpole}) vanishes,  and then it is not necessary to perform any regularization. However, from the last term of eq. (\ref{renorsusy}) it is possible to verify that the supersymmetric kink mass at one loop order is divergent at $B=\sqrt{2}$. This is a non-intuitive and unexpected  result since
it is commonly believed that supersymmetry improves, rather than spoils, the ultraviolet divergences in the theory. We believe that this fact deserves a better analysis, and to do that it would be interesting to examine the two-loop quantum corrections to this exotic supersymmetric model as it has been already done for the sine-Gordon and $\phi^4$ models \cite{Nastase,Vega}. This issue is an interesting subject for future investigations.



\vskip 1cm
\noindent
{\bf Acknowledgements} \\
\vskip .1cm \noindent
{The authors would like to thank J. A. Helayel Neto for reading the manuscript and for useful comments. The authors also thank the referees for useful comments and suggestions that helped us to improve the readability of our work, as well as to clarify some important issues.  
}


\end{document}